\documentclass{IEEEtran4PSCC}

\usepackage{myStyleIEEE}
\usepackage{nomencl}
\IEEEoverridecommandlockouts

\usepackage{etoolbox}
\renewcommand\nomgroup[1]{%
  \item[\bfseries
  \ifstrequal{#1}{P}{A. Parameters}{%
  \ifstrequal{#1}{V}{C. Variables}{%
  \ifstrequal{#1}{S}{B. Sets and Indices}{}}}%
]}

\makeatletter
\let\old@ps@headings\ps@headings
\let\old@ps@IEEEtitlepagestyle\ps@IEEEtitlepagestyle
\def\psccfooter#1{%
    \def\ps@headings{%
        \old@ps@headings%
        \def\@oddfoot{\strut\hfill#1\hfill\strut}%
        \def\@evenfoot{\strut\hfill#1\hfill\strut}%
    }%
    \def\ps@IEEEtitlepagestyle{%
        \old@ps@IEEEtitlepagestyle%
        \def\@oddfoot{\strut\hfill#1\hfill\strut}%
        \def\@evenfoot{\strut\hfill#1\hfill\strut}%
    }%
    \ps@headings%
}
\makeatother

\psccfooter{%
        \parbox{\textwidth}{\hrulefill \\ \small{23rd Power Systems Computation Conference} \hfill \begin{minipage}{0.2\textwidth}\centering \vspace*{4pt} \includegraphics[scale=0.06]{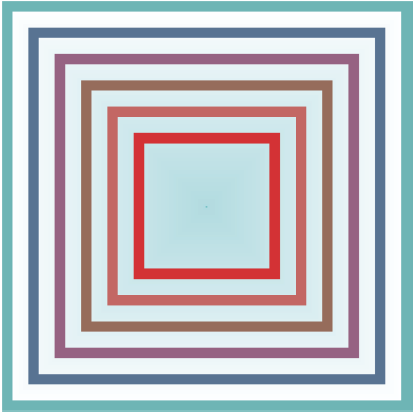}\\\small{PSCC 2024} \end{minipage} \hfill \small{Paris, France --- June 4 -- 7, 2024}}%
}

\begin{document}
\bstctlcite{IEEEexample:BSTcontrol}

\title{Pricing of Short Circuit Current in High IBR-Penetrated System}

\newtheorem{proposition}{Proposition}
\renewcommand{\theenumi}{\alph{enumi}}

\author{
\IEEEauthorblockN{Zhongda Chu, Jingyi Wu, Fei Teng}\\
\IEEEauthorblockA{Department of Electrical and Electronic Engineering \\
Imperial College London, London, UK\\
\{z.chu18, jingyi.wu19, fei.teng\}@imperial.ac.uk}
}

        
        
\maketitle
\IEEEpeerreviewmaketitle

\begin{abstract}
With the growing penetration of Inverter-Based Resources (IBRs) in power systems, stability service markets have emerged to incentivize technologies that ensure power system stability and reliability. Among the various challenges faced in power system operation and stability, a prominent issue raised from the increasing integration of large-scale IBRs is the significant reduction of the Short-Circuit Current (SCC) level in the system, which poses a considerable threat to system voltage stability and protection. Thus, a proper market mechanism to incentivize the provision of SCC as a stability service is desired. However, the pricing of this service within the future stability market has not yet been fully developed, due to the nonconvex nature of SCC constraints and the locational property of SCC. To address these problems, this work aims to explore, for the first time, a pricing model for SCC service by incorporating a linearized SCC constraint into the Unit Commitment (UC) problem, to achieve the desired SCC level and extract the shadow price for SCC through different pricing methods. 
\end{abstract}


\section{Introduction}
Growing awareness and acknowledgment of environmental issues like global warming and escalating pollution have motivated policymakers worldwide to commit to ambitious decarbonization targets \cite{29}. In recent years, there has been a profound transformation in the composition of the power generation mix. Particularly in Europe, electric systems are undergoing a shift towards a Renewable Energy Sources (RESs) dominated architecture\cite{30} \cite{28}. 

Interfaced with the grid through power electronic devices, IBRs provide much less SCC compared with traditional Synchronous Generators (SGs) due to their limited overloading capability \cite{10}. Therefore, with the increasing penetration of IBRs and the retirements of SGs, a decreasing trend of SCC levels in power systems is anticipated. CIRED Workshop predicts a reduction in SCC level for rural nodes in Germany \cite{26}. Also, according to System Operability Framework Document by National Grid ESO, an overall decline of around 15\% in the national SCC level from 2019 to 2023 is expected \cite{NGESO_1}. 

The reduction of SCC levels in power systems leads to threads in various aspects. Protection devices may take longer or even not be able to detect faults in the system, posing risks to system operation and stability \cite{AUEMO}. Additionally, inadequacy in SCC may cause unsettled voltage oscillations in the network and would further cause loss of synchronization of the Phase-Locked Loop (PLL) \cite{NGESO_2} and transient voltage stability issues. Therefore, enhancing the SCC and addressing the related issues have been an important procedure in high IBR-penetrated systems. Various attempts have been proposed on this, including physically improving the overloading capabilities of semiconductors by inserting phase-change material into chips \cite{44}, and allocating additional synchronous condensers in the system to maintain the system minimum SCC \cite{Jia18}. However, these approaches bring additional cost to either IBR manufacturers or system operators, and therefore how to incentivize the provision of SCC in the system becomes an essential element for the future SCC market development.

European Union Emissions Trading Scheme (EUETS) classifies the typical ancillary services procured by Transmission System Operators (TSOs) into frequency control services (regulation, load following, operating reserves) and non-frequency control services (voltage control, ramp limit, SCC, etc.) \cite{33}, which are considered as the main ancillary services in existing electricity markets \cite{27}. Although some of these markets have been well established based on duality theory, difficulties still exist. In order to determine the price of SCC in the system, it is necessary to integrate the SCC constraint into the system scheduling model. However, due to the complicated SCC calculation, proper linearization methods such as \cite{9329077,9} to relax the nonlinearity and nonconvexity in the SCC constraint have to be incorporated. 

Another key challenge lies in determining the marginal costs of uplift payment, which is associated with the inherent nonconvexity in power systems: the binary nature of the commitment decision for thermal units. This makes it impractical to apply duality theory directly, even when pricing energy in energy markets \cite{3}. Some approaches have been proposed to address this issue. The dispatchable method relaxes the commitment decisions of thermal units from binary variables to continuous ones. This approach converts the original MILP formulation of the UC problem into an LP problem where the dual variables can be defined properly \cite{48}. Nevertheless, it has a limitation in capturing the complete technical characteristics of generators, therefore prices computed through this method might not correspond to feasible operating points for power systems \cite{2}. The restricted pricing method, on the other hand, computes the SCC price by solving the UC problem twice. In the first time, the original MILP is solved as usual to obtain the optimal commitment decisions; Subsequently, a second optimization is performed by relaxing the binary variables while employing equality constraints to enforce the commitment variables to retain their optimal values. Thereby the operating conditions of the power system can be accurately represented when determining the dual variables. However, this approach adds a new term to the Lagrangian function, where a ``price for commitment'' is thus created, leading to a side payment for units associated with a commitment decision variable \cite{zhang2009reducing}. The concept of marginal unit price is introduced in \cite{mukherjee2006average} to determine the prices in integer programs, where the right-hand side resource availability can only be varied in a discrete manner. However, none of the above methods have been studied in the SCC market setting. 

\textcolor{black}{Moreover, even with proper relaxation of the nonconvex SCC constraints, applying the duality theory only gives the SCC price at each sink (nodes requiring SCC) and it is not clear how this price should be allocated to different SCC sources (units contributing SCC).} There is no research that has looked into this problem. In this context, to incentivize system components such as conventional SGs and IBRs to provide sufficient SCC at critical locations, the pricing schemes of SCC are investigated and analyzed. Our main contributions include the following:
\begin{itemize}
    \item A novel SCC pricing scheme is developed by incorporating linearized SCC constraints into the system UC problem such that the contribution to the SCC from different devices in the system can be properly quantified.
    \item The nonconvexity due to the existence of binary variables is dealt with by applying three different methods, namely the dispatchable method, the restricted method, and the marginal unit price method. 
    \item The prices obtained by different schemes are compared, and the impact of system conditions on the SCC price is analyzed in the modified IEEE 39-bus system with valuable information obtained for future SCC market development.
\end{itemize}
The rest of this paper is structured as follows. Section~\ref{sec:2} introduces the SCC constraint and its linearization approach. The SCC pricing schemes based on the SCC-constrained UC problem are developed in Section~\ref{sec:3}, followed by case studies in Section~\ref{sec:4}. Finally, Section~\ref{sec:5} concludes the paper and discusses the future work.

\section{SCC Constraint Representation and Linearization} \label{sec:2}
Consider a general multi-machine system with SGs $g\in\mathcal{G}$ and IBRs $c\in \mathcal{C}$. By modeling the IBR as current sources, the SCC constraint  $I_{sc_F}^{''}\ge I_{F_{\mathrm{lim}}}^{''}$ in a high IBR-penetrated system can be derived based on the superposition principle \cite{9329077}:
\begin{subequations}
\label{SCC_C}
\begin{align}
    I_{sc_F}^{''} & = \frac{-\sum_{g\in\mathcal{G}}Z_{F\Psi(g)} I_{g} x_g-\sum_{c\in \mathcal{C}} Z_{F\Phi(c)}I_{c} \alpha_c } {Z_{FF}} \ge I_{F_{\mathrm{lim}}}^{''} \label{SCC_constr} \\
    Z & = Y^{-1} \label{ZY}
\end{align}
\end{subequations}
where $I_{sc_F}^{''}$ and $I_{F_{\mathrm{lim}}}^{''}$ are the short circuit current at bus $F$ and the corresponding limit; $\Psi(g)/\Phi(c)$ maps $g\in \mathcal{G}/c\in \mathcal{C}$ to the corresponding bus index; $I_g/I_{c}$ is the SCC contribution from SG $g$/IBR $c$; $x_g$ is the binary variable representing the SG online status; $\alpha_c$ is the online capacity in percentage of IBR $c$; $\mathbb{I}_N$ is $N$-dimensional identity matrix; $Z/Y$ is the system impedance/admittance matrix, defined as follows:
\begin{equation}
    \label{Y}
    Y = Y^0 +  Y^g,
\end{equation}
where $Y^0$ is the admittance matrix of the transmission lines only; $Y^g$ denotes the additional $Y$ matrix increment due to SGs' subtransient reactance. Depending on the state of SGs, the elements in $Y^g$ can be expressed as:
\begin{equation}
\label{Y2}
    Y_{ij}^g=
    \begin{cases}
    \frac{1}{X_{dg}^{''}}x_g\;\;&\mathrm{if}\,i = j \land \exists\, g\in \mathcal{G},\, \mathrm{s.t.}\,i=\Psi(g)\\
    0\;\;& \mathrm{otherwise}.
    \end{cases}
\end{equation}

It is understandable that due to the dependence of $Y$ on $x_g$, the matrix inverse with binary decision variables in \eqref{ZY} is in general difficult to be included in an optimization problem, which is essential for SCC pricing. The data-driven method proposed in \cite{9329077} is adapted to linearize the SCC constraint, which is briefly discussed as follows. Introduce a set of new parameters $\mathcal{K} = \left\{k_{Fg}, \; k_{Fm}, \; k_{Fc} \right\},\,\forall F,g,m,c$  as the coefficients in the linearized SCC constraint, which reformulates \eqref{SCC_C} into:
\begin{align}
\label{SCC_Linear}
I_{{F}_{L}}^{''} = \sum_{g \in \mathcal{G}} k_{Fg} x_{g} + \sum_{c \in C} k_{Fc} \alpha_{c} + \sum_{m \in M} k_{Fm} \eta_{m} \geq I_{{F}_{\mathrm{lim}}}^{''},
\end{align}
where $I_{{F}_{L}}^{''}$ is the linearized SCC. The interactions between every pair of SGs are captured by the second-order term \(\eta_m\) to accommodate the nonlinearity in \eqref{SCC_C}, which is defined as:
\begin{subequations}
\begin{align}
    \label{eq:x1x2}
    \eta_m  &=x_{g_1}x_{g_2},\quad \mathrm{s.t.}\{g_1,\,g_2\}=m\\
    m\in\mathcal{M} & =\{g_1,\,g_2 \mid g_1,\,g_2\in \mathcal{G}\}.
\end{align}
\end{subequations}
The product of two binary variables can be further linearized by:
\begin{subequations}
\begin{align}
\eta_m & \leq x_{g_1}\\
\eta_m & \leq x_{g_2}\\
\eta_m & \geq x_{g_1}+x_{g_2}-1.
\end{align}
\end{subequations}
The parameters in $\mathcal{K}$ are determined by solving the following optimization problem:
\begin{subequations}
\label{DM3}
\begin{align}
    \label{obj3}
    \min_{\mathcal{K}}\quad & \sum_{\omega \in \Omega_2} \left(I_{sc_F}^{''(\omega)} - I_{F_L}^{''(\omega)} \right)^2\\
    \label{coef_ctr2}
    \mathrm{s.t.}\quad & I_{F_L}^{''(\omega)}< I_{F_{\mathrm{lim}}}^{''},\,\,\forall \omega \in \Omega_1\\
    \label{coef_ctr3}
    & I_{F_L}^{''(\omega)}\ge I_{F_{\mathrm{lim}}}^{''},\,\,\forall \omega \in \Omega_3
\end{align}
\end{subequations}
with $(\cdot)^{(\omega)}$ denoting quantities associated with sample $\omega$, and $\omega = \{x_g^{(\omega)},\,\alpha_c^{(\omega)},\, I_{sc_F}^{''(\omega)}\}\in \Omega$ denoting the data set. It is generated by evaluating the SCC at each bus in representative system conditions. $\Omega_1 ,\, \Omega_2$ and $\Omega_3 $ are the subsets of $\Omega$, being defined as follows:
\begin{subequations}
\begin{align}
    \Omega &= \Omega_1 \cup\Omega_2\cup\Omega_3 \\
    \label{Omega1}
    \Omega_1 & = \left\{\omega\in \Omega \mid I_{sc_F}^{''(\omega)}<I_{F_{\mathrm{lim}}}^{''} \right\}\\
    \label{Omega2}
    \Omega_2 & = \left\{\omega\in \Omega \mid I_{F_{\mathrm{lim}}}^{''} \le I_{sc_F}^{''(\omega)}<I_{F_{\mathrm{lim}}}^{''} + \nu \right\}\\
    \label{Omega3}
    \Omega_3 & = \left\{\omega\in \Omega \mid I_{F_{\mathrm{lim}}}^{''} + \nu\le I_{sc_F}^{''(\omega)} \right\}.
\end{align}
\end{subequations}
Given \eqref{coef_ctr2} and \eqref{Omega1}, all the below-limit SCC can be classified correctly by the linearized function. Ideally, it is also desired to identify all the above-limit SCC. However, this may cause infeasibility due to the restricted linear structure in \eqref{SCC_Linear}. Therefore, a parameter $\nu\in \mathbb{R}^+$ is introduced to define $\Omega_2$ and $\Omega_3$ as in \eqref{Omega2} and \eqref{Omega3}. In this way, all the data points in $\Omega_3$ will be classified correctly and misclassification can only occur in $\Omega_2$. Furthermore, $\nu$ should be chosen as small as possible while ensuring feasibility.

\section{Pricing SCC Services}  \label{sec:3}
\textcolor{black}{In order to better explain the concept of SCC price, we first define two sets of nodes, i) SCC sources ($\mathcal{S}_1$): all the nodes that provide SCC to other nodes during faults, i.e., SG and IBR nodes and ii) SCC sinks ($\mathcal{S}_2$): all the nodes with SCC shortage. It is therefore desired to obtain the SCC price from each source to each sink, leading to an overall price matrix $p_{SCC}\in\mathbb R^{|\mathcal{S}_1|}\times \mathbb R^{|\mathcal{S}_2|}$. To derive the SCC prices, the SCC-constrained UC model is formulated as follows.}

\subsection{SCC-constrained UC} \label{sec:3.1}
The objective of the UC problem is to minimize the expected cost over all nodes in the given scenario tree:
\begin{equation}
    \label{eq:SUC}
    \min \sum_{n\in \mathcal{N}} \pi (n) \left( \sum_{g\in \mathcal{G}}  C_g(n) + \Delta t(n) c^s P^s(n) \right),
\end{equation}
where $\pi(n)$ is the probability of scenario $n\in \mathcal{N}$ and $C_g(n)$ is the operation cost of unit $g\in \mathcal{G}$ in scenario n including startup, no-load and marginal cost; $\Delta t(n)c^sP^s(n)$ represents the cost of the load shedding in scenario n with the three terms being the time step of scenario n, load shedding cost and shed load. The objective function \eqref{eq:SUC} is subjected to a number of constraints, including the linearized SCC constraint \eqref{SCC_Linear}. All other conventional UC constraints such as those related to power balance, thermal units and transmission lines are not listed in the paper. The readers can refer to \cite{chu2023stabilityII} for details.

However, even with the linearized SCC constraints, the binary nature of the commitment decision for thermal units makes it impractical to apply the duality theory for SCC pricing. Hence, three pricing schemes, namely the dispatchable method, the restricted method, and the marginal unit price method, are investigated to cope with the non-continuity in the short-circuit current constraint. With relaxed and convex SCC constraints, the duality theory can be applied to extract the SCC prices in the system. The pros and cons of each method and their applying conditions are discussed in the following sections.

\subsection{Pricing schemes for SCC service }
Note that the main issue to be addressed in order to extract a meaningful SCC dual variable is the discrete online/offline characteristic of SGs and the fact that they do not provide SCC in a continuous manner. Regarding this problem, three alternative pricing methods have been proposed in this section.

\subsubsection{Dispatchable Pricing}
This approach involves direct linear relaxation of the binary decision variables $x_g$, i.e replacing the original constraint $x_g\in\{0,1\}$ with
\begin{equation}
\label{binary_relx}
0 \leq x_{g} \leq 1,\,\forall g\in \mathcal{G}.
\end{equation}
However, this causes the quadratic term \(\eta_m = x_{g_1}\cdot x_{g_2}\) in \eqref{SCC_Linear} now to become a product of two continuous decision variables and can no longer be linearized with the big-M method. Some attempts have been proposed to deal with this type of bilinear product terms. Reference \cite{56} proposes a piecewise linear method which divides the quadratic term into multiple segments that are approximated by straight lines. However, additional binary decision variables must be introduced to ensure proper segmentation and weighing of each fragment for the closest approximation, which reproduces nonconvexity in the problem. Therefore, in order to apply the dispatchable pricing method the term $\eta_m$ in constraint \eqref{SCC_Linear} is removed, resulting in the SCC constraint with the form below:
\begin{align}
\label{SCC_Linear2}
I_{{F}_{L}}^{''} = \sum_{g \in \mathcal{G}} k_{Fg} x_{g} + \sum_{c \in C} k_{Fc} \alpha_{c} \geq I_{{F}_{\mathrm{lim}}}^{''}.
\end{align}
This simplification incurs the drawback of sacrificing some generality for this model, where the extracted price might not align with a viable operating point for the system. Nevertheless, the price of the SCC as a service can be obtained from the Karush Kuhn Tucker (KKT) conditions of the SCC-constrained optimization problem. Since the SCC constraints are linear, the price is immediately obtained from the dual variables of the constraints. The terms associated with SCC constraints and their corresponding Lagrange multipliers take the form:
\begin{align} \label{dual_SCC}
     -\lambda_{SCC} \left( \sum_{g \in \mathcal{G}} k_{Fg} x_{g} + \sum_{c \in C} k_{Fc} \alpha_{c} - I_{{F}_{\mathrm{lim}}}^{''} \right).
\end{align}
\textcolor{black}{
The shadow price for SCC at sink $F\in\mathcal{S}_2$ with contribution from all the sources can then be obtained by evaluating the KKT stationary condition, which is given by differentiating the Lagrangian with respect to
$I_{{F}_{\mathrm{lim}}}^{''}$: 
\begin{equation} 
    p_{SCC}^{\varsigma F} = \frac{\mathrm{d} \, \mathcal{L}}{\mathrm{d} \, I_{{F}_{\mathrm{lim}}}^{''}} = \lambda_{SCC}.
\end{equation}
The price from a single SCC source $E\in\mathcal{S}_1$ to the sink $F\in\mathcal{S}_2$ can then be calculated by considering its weighted contribution as indicated in \eqref{SCC_C}:
\begin{equation} 
    p_{SCC}^{E\rightarrow F} = \frac{Z_{FE} I_E}{\sum_{E\in\mathcal{S}_1} Z_{FE} I_E} p_{SCC}^{\varsigma F}, 
\end{equation}
where $I_E$ is the equivalent current injection, i.e., $I_g x_g$ in \eqref{SCC_C}.} However, as discussed in the introduction, relaxation of the binary variables in \eqref{binary_relx}, may lead to infeasible operating conditions for the system. To address this issue, the following two methods are considered. 

\subsubsection{Restricted Pricing}
In the restricted model, the original MILP-based UC problem is solved first, with the optimal decisions regarding the binary decision variables \(x_g\) being recorded. Then, the optimization problem is solved again with the binary variables $x_g$ being set to their optimal values \(x_g^*\) based on the solution of the previous problem. Therefore, in the second problem, constraints that confine binary variables are replaced with:
\begin{subequations} \label{xg*}
\begin{align}
x_{g} &= x_{g}^\ast \\
\eta_{m} &= \eta_{m}^\ast. 
\end{align}
\end{subequations}
Subsequently, the terms associated with SCC constraints and their corresponding Lagrange multipliers can be expressed:
\begin{align} \label{dual_SCC2}
     -&\lambda_{SCC} \left( \sum_{g \in \mathcal{G}} k_{Fg} x_{g} + \sum_{c \in C} k_{Fc} \alpha_{c} + \sum_{m \in M} k_{Fm} \eta_{m} - I_{{F}_{\mathrm{lim}}}^{''} \right) \nonumber\\
     + & \lambda_{commit} \left( x_g^\ast - x_g \right).
\end{align}
It should be noted that with this method, the product terms $\eta_m$ no longer need to be neglected as in \eqref{SCC_Linear2}. Instead, \eqref{SCC_Linear} can be used in the UC problem, leading to the expression in \eqref{dual_SCC2}.

As a result, two prices can be derived accordingly. For the SCC price, the same result as the dispatchable method holds, i.e.,
\begin{equation} 
    \frac{\mathrm{d} \, \mathcal{L}}{\mathrm{d} \, I_{{F}_{\mathrm{lim}}}^{''}} = \lambda_{SCC}.
\end{equation}
However, due to the discontinuity nature of the SG online status, forcing $x_g$ to be their optimal value by introducing \eqref{binary_relx} is very likely to make the SCC constraint non-binding, which causes $\lambda_{SCC}$ (the SCC price) to be zero. To address this issue, another price is defined here that reflects the value an SG being committed online, which can be derived by taking the derivative of the Lagrange function with respect to $x_g^*$:
\begin{equation} 
\label{p2}
    \frac{\mathrm{d} \, \mathcal{L}}{\mathrm{d} \, x_g^*} = \lambda_{commit}.
\end{equation}
This price has a different interpretation from the SCC price. The latter represents the increment of the system operation cost (the price the system operator is willing to pay) if the SCC requirement is increased by $1\, \mathrm{p.u.}$, whereas in this case the price is defined for each generator $g\in\mathcal{G}$ and reflects the variation in system operation cost due to the change in the SGs' commitment status. 

However, as mentioned earlier this approach has several pitfalls. An implication is the introduction of a side payment for the committed units, which can potentially result in insufficient revenue for the system operator. Furthermore, the extracted shadow price \(\lambda_{commit}\) can exhibit high volatility between large positive numbers and zero, leading to frequent instances of seemingly unfair outcomes. Additionally, the unit commitment price defined by \eqref{p2} contains not only the cost to maintain the SCC constraints, since the unit commitment decisions of SGs are also coupled with other constraints that influence the system operation cost, such as power balance constraints. Therefore, the price obtained in this way could potentially be much higher than the actual price of the SCC service. To decouple the impact of the SG commitment decision on the SCC from other constraints in the system, an alternative marginal unit pricing scheme is discussed in the next subsection.

\subsubsection{Marginal Unit Pricing}
This approach shares a similar physical interpretation to the restricted pricing model but does not require the application of the duality theory. Therefore, it can deal with the situation where the marginal price defined by the gradient of the Lagrange function does not exist due to the nonconvex nature of the resources. The marginal unit price quantifies the economic value of the SCC contribution provided by different generation units. It is obtained by solving the system scheduling model, where a commitment price is defined for each SG according to its contribution to system SCC.  

First, solve the original optimization problem defined in Section~\ref{sec:3.1}, which is denoted by the following compact form:
\begin{subequations}
\label{UC_compact}
    \begin{align}
        \min_{\mathsf{X}}\;\; &\mathsf{f} (\mathsf{X})\\
        \mathrm{s.t.}\;\; & \mathsf K^{\mathsf T} \mathsf X \ge I_{{F}_{\mathrm{lim}}}^{''} \label{SCC_MUP} \\
        & h(\mathsf{X})\ge 0,
    \end{align}
\end{subequations}
where $\mathsf X$ is the decision variable; $\mathsf f(\mathsf{X})$ is the objective function; \eqref{SCC_MUP} is the system SCC constraint and $h(\mathsf{X})$ is the rest of the constraints in the UC problem. Denote the optimal value of problem \eqref{UC_compact} as $\mathsf{f}^*$. To determine the economic value of an SCC source $E\in\mathcal{S}_1$, the following problem is solved.
\begin{subequations}
\label{UC_delta}
    \begin{align}
        \min_{\mathsf{X}}\;\; &\mathsf{f} (\mathsf{X}) \label{f_X2}\\
        \mathrm{s.t.}\;\; & (\mathsf K + \Delta \mathsf K^E)^{\mathsf T} \mathsf X \ge I_{{F}_{\mathrm{lim}}}^{''} \label{Delta_g}\\
        & h(\mathsf{X})\ge 0 \label{k_X2}.
    \end{align}
\end{subequations}
The vector $\Delta \mathsf K^E$ and is of the same dimension as $\mathsf K$, which is defined as:
\begin{align}
    \Delta \mathsf K^E_i =  &
    \begin{cases}
        0\qquad\quad\; ,\mathrm{if}\, i\neq E\\
        -\mathsf{K}_{i}\qquad,\mathrm{if}\,i = E\\
    \end{cases}
\end{align}
where $(\cdot)_{i}$ denotes the $i$-th element of the vector. With the above definition, the coefficient in $\mathsf K + \Delta \mathsf K^E$ corresponding to the $E$-th element in $\mathsf{X}$, becomes zero, thus eliminating the SCC provision from the SCC source $E$, whereas its impacts on the operation cost \eqref{f_X2} and other constraints \eqref{k_X2} remain the same. \textcolor{black}{Denoting the optimal value of problem $\eqref{UC_delta}$ as $\mathsf{f}^{*}_E$ enables us to define the unit price of the source $E\in\mathcal{S}_1$ with contribution to the sink $F\in\mathcal{S}_2$:
\begin{equation}
\label{mu_g}
    p_{unit}^{E\rightarrow F} = \mathsf{f}^{*}_E-\mathsf{f}^{*}.
\end{equation}
}

\section{Results} \label{sec:4}
In order to investigate the value of SCC provision from different resources, and how it is influenced by system operation, we consider an optimization problem with a time horizon of 24 hours and a time step of 1 hour, which is implemented in Matlab with the YALMIP toolbox \cite{24} and solved by Gurobi (10.0.0) on a PC with Intel(R) Core(TM) i7-10750H CPU @ 2.60GHz and RAM of 16 GB. The weather conditions are obtained from online numerical weather prediction \cite{weather}. 

\begin{figure}[!t]
    \centering
    \vspace{-0.4cm}
	\scalebox{0.5}{\includegraphics[trim=0.3cm 0 1.4cm 0,clip]{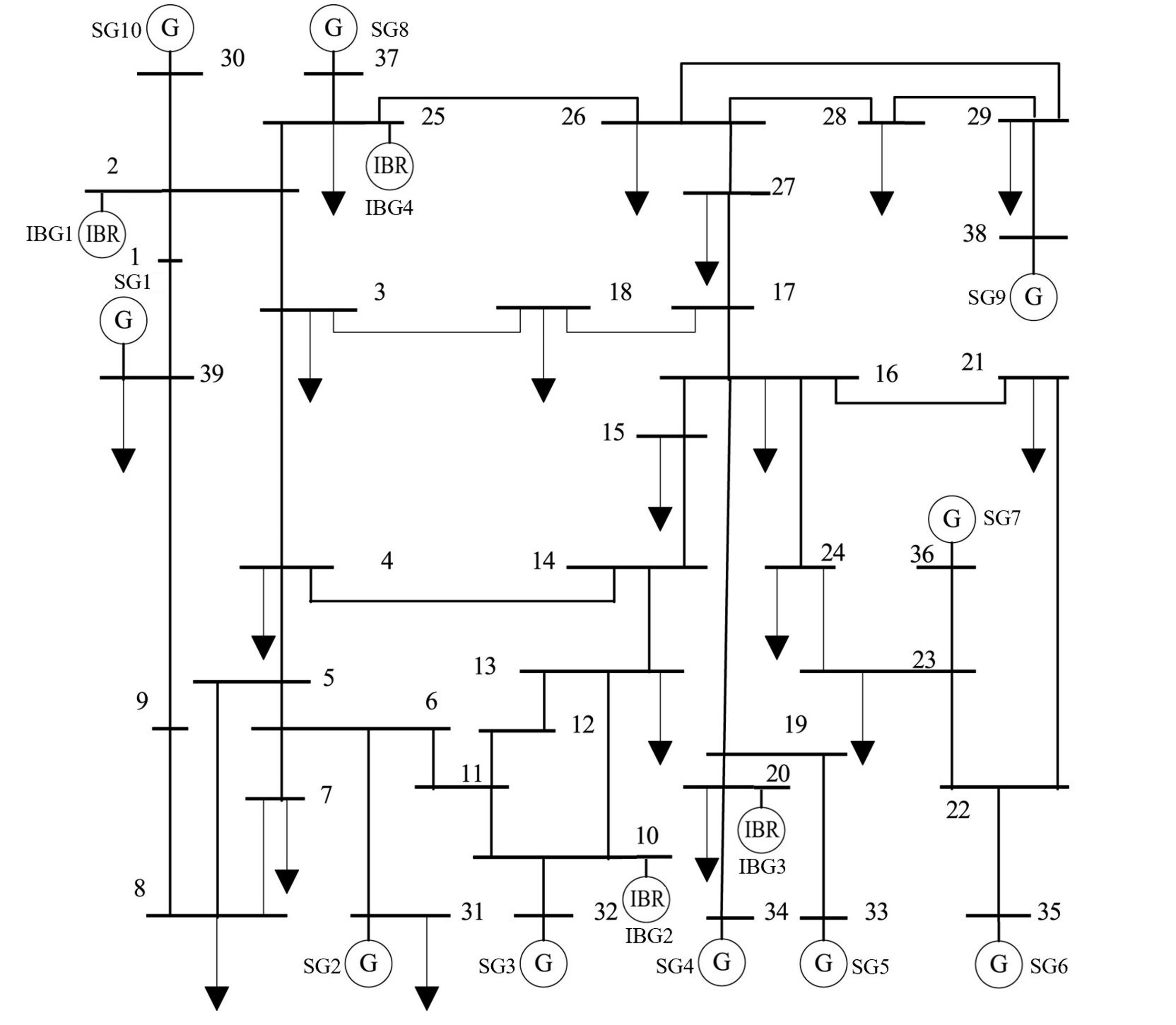}}
    \caption{\label{fig:39-bus}Modified IEEE-39 bus system.}
    \vspace{-0.4cm}
\end{figure}

The simulation is carried out based on the modified IEEE 39-bus system to illustrate the validity of the proposed SCC linearization models and verify the feasibility of the pricing schemes. The topology of this system is as shown in Fig.~\ref{fig:39-bus}, which consists of 10 synchronous generators (SCC sources), 39 buses, 46 lines and 4 IBRs (SCC sinks). Some system parameters are set as follows: load demand $P^D\in [5.16, 6.24]\,\mathrm{GW}$, base power $S_B = 100\mathrm{MVA}$. Other parameters can be found in \cite{4113518,chu2023stabilityII}.

\subsection{SCC Constraint Validation}
\begin{figure}[!t]
    \centering
    \vspace{-0.4cm}
    \includegraphics[width=\columnwidth,keepaspectratio,trim=4cm 9.7cm 4.5cm 10cm,clip]{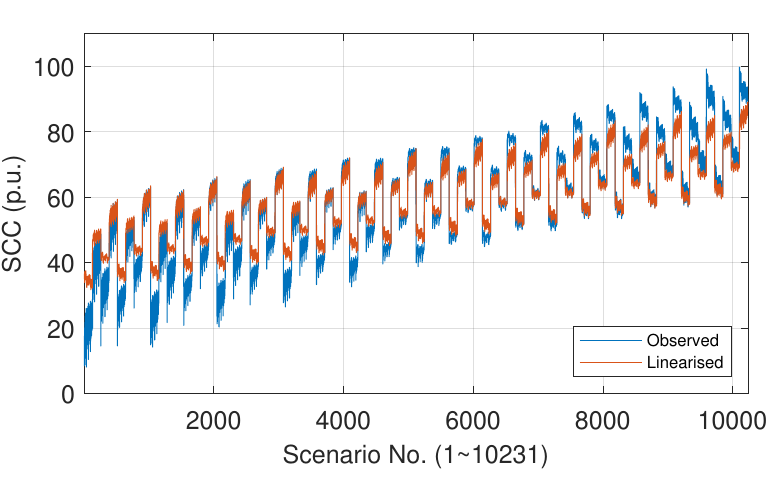}
    \caption{Observed and linearized SCC for IBG buses (margin $\nu$ = 1)}
    \label{fig6.1} 
    \vspace{-0.4cm}
\end{figure}

To obtain an accurate approximation of the SCC constraint, the number of wind conditions considered when training the fitting model \eqref{DM3}, should be large enough so that the mapping defined in \eqref{SCC_Linear} can effectively cover a wide range of real-world weather conditions. In this particular case study,  \(\mathcal{K}\) is defined with respect to all combinations of the SG's commitment statuses with ten different wind capacity values, ranging from $0$ to $0.9\,\mathrm{p.u.}$ with a step of $0.1\,\mathrm{p.u.}$, which gives a total of \((2^{10}-1) \times 10 = 10230\) different scenarios. We exclude the scenario where all SGs are offline, as it is not realistic considering the system demand throughout the day. Having obtained \(\mathcal{K}\), the observed \(I_{{sc}_{F}}^{''}\) and linearized \(I_{F_{L}}^{''}\) can be aligned for comparison, as illustrated in Fig.~\ref{fig6.1}. Note that only the SCC at IBR~1 bus is shown as a similar trend can be observed for the others.

In order to examine the performance of the linearization method, two types of errors are analyzed with Type I being the SCC which satisfies the constraint according
to the linearized value but actually violates and Type II being the opposite:
\begin{subequations}
\begin{align}
\text{Type I} : I_{F_{L}}^{''} \geq I_{F_{\mathrm{lim}}} & \cap \, I_{{sc}_{F}}^{''} < I_{F_{\mathrm{lim}}}\\
\text{Type II} : I_{F_{L}}^{''} < I_{F_{\mathrm{lim}}} & \cap \, I_{{sc}_{F}}^{''} \geq I_{F_{\mathrm{lim}}}.
\end{align}
\end{subequations}
In terms of the actual consequences for power systems, the occurrence of Type I errors is much more severe than Type II errors, as they may lead to insecure operating conditions, thus being undesired. Each type of error is quantified by their total number of occurrences \(N_e\) and the averaged relative value \(err\), defined as:
\begin{equation} \label{6.3.3}
N_e = |\mathcal{E}|
\end{equation}
\begin{equation}\label{6.3.4}
err = \frac{1}{N_e} \sum_{\mathcal{E}}\frac{I_{F_{L}}^{''(\mathcal{E})}-I_{{sc}_{F}}^{''(\mathcal{E})}}{I_{{sc}_{F}}^{''(\mathcal{E})}},
\end{equation}
in which \(\mathcal{E}\) denotes the set of errors. Results evaluated at different IBR buses are presented in Table~\ref{tab:error}. As expected, the SCC approximation completely eliminates Type I error and produces only a small amount of Type II errors, hence presenting a good performance.

\begin{table}[!b]
\renewcommand{\arraystretch}{1.2}
\vspace{-0.4cm}
\caption{SCC Linearization Errors at IBR Buses}
\label{tab:error}
\noindent
\centering
    \begin{minipage}{\linewidth} 
    \renewcommand\footnoterule{\vspace*{-5pt}} 
    \begin{center}
        \begin{tabular}{ c | c | c | c | c  }
            \toprule
             \multirow{2}{4em}{\textbf{Location}} & \multicolumn{2}{c|}{\textbf{Type I Error}} &\multicolumn{2}{c}{\textbf{Type II Error}}  \\ 
            \cline{2-5}
            &$N_e$ & $err$ & $N_e$ & $err$ \\ 
            \cline{1-5} 
            IBR 1 & $0$ & $0$ & $609$ & $-0.55\%$ \\
            \cline{1-5} 
            IBR 2 & $0$ & $0$ & $624$ & $-0.63\%$ \\
            \cline{1-5} 
            IBR 3 & $0$ & $0$ & $687$ & $-0.89\%$ \\
            \cline{1-5} 
            IBR 4 & $0$ & $0$ & $666$ & $-0.79\%$ \\
           \bottomrule
        \end{tabular}
        \end{center}
    \end{minipage}
    \vspace{-0.4cm}
\end{table} 

\begin{figure}[!t]
    \centering    \includegraphics[width=\columnwidth,keepaspectratio,trim=4cm 9.7cm 4.5cm 10cm,clip]{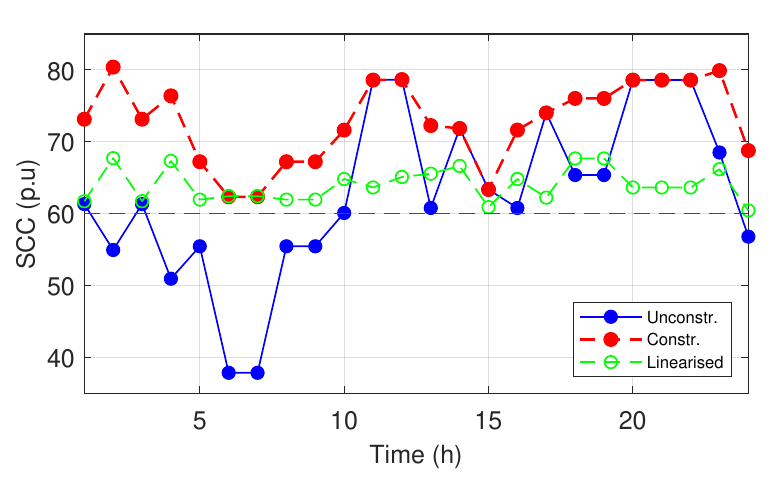}
    \caption{SCC at the bus of IBR 1: constrained vs unconstrained.}
    \label{fig6.2}
    \vspace{-0.4cm}
\end{figure}

To demonstrate the effectiveness of the implemented SCC constraints, the actual SCC before and after adding the constraint at the IBR bus, represented by ``Uncontr.'' and ``Constr." respectively is depicted in Fig.~\ref{fig6.2}. Note that due to the similarity, only the situation in IBR 1 is shown. It can be clearly identified that once the constraint is enforced, all SCC values that are initially below the limit in the unconstrained situation are now raised to meet the constraint. Also, it is worth noticing that the overlapping parts of the Constr. and Unconstr. lines indicate that the unconstrained current at this period of time is originally well above the limit, thus remaining unchanged. Furthermore, it is evident that the linearized values do not align with the actual SCC values for most of the hours in the constrained case. However, as long as the actual SCC and the linearized one are on the same side of the limit, the error is of no concern, as explained in Section~\ref{sec:2}.

\subsection{SCC Pricing} 
The SCC prices obtained with different methods are demonstrated in this section. For simplicity, only one sink is active in each UC problem.
\subsubsection{Dispatchable Method}
\begin{figure}[!b]
    \vspace{-0.4cm}
    \centering    \includegraphics[width=\columnwidth,keepaspectratio,trim=4cm 9.7cm 4.5cm 10cm,clip]{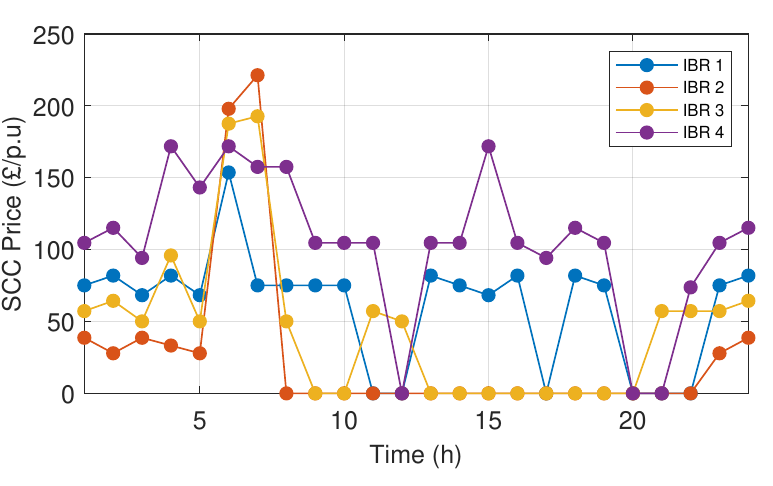}
    \caption{\label{fig:price_disp}SCC prices at IBR buses during one-day operation.}
    \vspace{-0.4cm}
\end{figure} 
The SCC prices at each hour for each sink $(p_{SCC}^{\varsigma F})$, obtained based on the dispatchable method are illustrated in Fig.~\ref{fig:price_disp}. On average, the SCC prices for the IBR 1, IBR 2, IBR 3, and IBR 4 buses are approximately {60.46 \pounds/h}, {27.18 \pounds/h}, {45.49 \pounds/h}, and {105.17 \pounds/h}, respectively. Furthermore, these prices vary at different IBR buses and hours during the day. The variations are mainly attributed to the differences between the unconstrained SCC and the limits at different IBR buses: buses with a greater need for SCC enhancement tend to have higher marginal costs. The spikes in prices are related to the start-up cost of the thermal units, as more SGs are needed for the SCC provision. In addition, the periods when the SCC price equals zero indicate that the SCC constraints during these hours are non-binding. This means that no ancillary service is required because the committed units, determined by other predominant constraints, have already provided sufficient SCC. \textcolor{black}{The payment from each SCC sinks to different sources can also be determined by evaluating $p_{SCC}^{E\rightarrow F} (Z_{FE}/Z_{FF}) I_E$, giving the results in Table~\ref{tab:prices}. In general, an SG with higher capacity and shorter electrical distance to the SCC sink contributes more SCC, thus being paid with more money, such as ${10\rightarrow 1}$ and ${3\rightarrow 2}$ (arrows indicating SCC direction).
}

\begin{table}[!t]
\renewcommand{\arraystretch}{1.2}
\caption{\textcolor{black}{SCC Payment from Different Sinks}}
\label{tab:prices}
\noindent
\centering
    \begin{minipage}{\linewidth} 
    \renewcommand\footnoterule{\vspace*{-5pt}} 
    \begin{center}
        \begin{tabular}{ c | c | c | c | c | c | c | c | c | c | c}
            \toprule
             \multirow{2}{*}{\textbf{Sink}} & \multicolumn{10}{c}{\textbf{Source (SGs) [$\mathrm{\times10^2\pounds/h}$]}} \\ 
            \cline{2-11}
            &\textbf{1} &\textbf{2} &\textbf{3} &\textbf{4} &\textbf{5} &\textbf{6} &\textbf{7} &\textbf{8} &\textbf{9} &\textbf{10}  \\
            \cline{1-11}
            IBR 1 & $3.3$ & $1.4$ & $0$ & $0$ & $0$ & $0$ & $0$ & $0$ & $0$ & $9.1$ \\
            \cline{1-11} 
            IBR 2 & $1.5$ & $1.6$ & $2.7$ & $0$ & $0$ & $0$ & $0$ & $0$ & $0$ & $0$ \\
            \cline{1-11} 
            IBR 3 & $0.6$ & $0.4$ & $0$ & $0.3$ & $7.8$ & $0$ & $0$ & $0$ & $0$ & $0$ \\
            \cline{1-11} 
            IBR 4 & $6.0$ & $2.6$ & $0$ & $0$ & $0$ & $0$ & $0$ & $1.9$ & $0$ & $0$ \\
           \bottomrule
        \end{tabular}
        \end{center}
    \end{minipage}
    \vspace{-0.5cm}
\end{table}

\subsubsection{Restricted Method}
As discussed earlier, the SCC price obtained with this method is zero for all hours. This is because forcing the SG commitment decisions to be their optimal value makes the SCC constraint non-binding, as the SGs provide the SCC in a discrete manner. Therefore, the prices for the unit commitment decisions are investigated. It is observed that the restricted method is not suitable for SCC pricing due to the extremely high prices up to $38\,\mathrm{k\pounds/h}$ (results not shown due to the space limitation). 

Possible reasons for this could be that the objective function is highly sensitive to changes in commitment status, which directly influence the no-load cost and start-up cost of SGs. Also, the generator states indirectly impact the operational cost through the power balance constraints, which may lead to a significant price due to the high penalty of load shedding. Therefore, the dual variables extracted with respect to the commitment decisions reflect not only the effect of SCC constraint variation, but also all other constraints correlated to \(x_g\). The binary nature of these variables makes the price highly volatile. It is not possible to isolate \(y_g\) in SCC constraint from all other relevant terms, hence this approach is not applicable to SCC pricing. 


\subsubsection{Marginal Unit Price Method}
\begin{figure}[!b]
    \centering
    \vspace{-0.4cm}\includegraphics[width=\columnwidth,keepaspectratio,trim=4cm 9.7cm 4.5cm 10cm,clip]{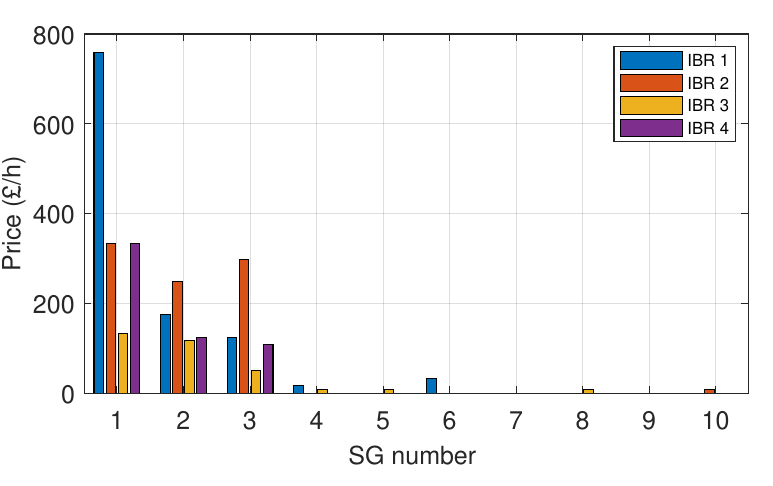}
    \caption{\label{fig:P3}SG commitment prices for different IBRs.}
    \vspace{-0.4cm}
\end{figure} 
Similar to the restricted method, the averaged prices of SGs during 24-hour operation at different IBR buses are depicted in Fig.~\ref{fig:P3}. It can be observed that the first three SGs have a significant commitment price for the SCC provision, whereas the commitment prices for other SGs are close to zero. This is due to the fact that the first three SGs have a larger capacity and lower marginal cost, and hence are dispatched online most of the time, providing most of the SCC in the system. Therefore, the values of SCC supplied from these units are much higher than the rest, since their not providing SCC renders more operation cost, induced by forcing other units with higher cost online. As for the differences between the commitment prices of the same unit for different IBRs, this is due to the fact that the contribution of SGs to the SCC at different IBR buses varies depending on the electric distance between them. As a result, the SG tends to have a higher price for the SCC enhancement at a closer bus.

\begin{figure}[!t]
    \centering   
    \vspace{-0.4cm}
    \includegraphics[width=\columnwidth,keepaspectratio,trim=4cm 9.7cm 4.5cm 10cm,clip]{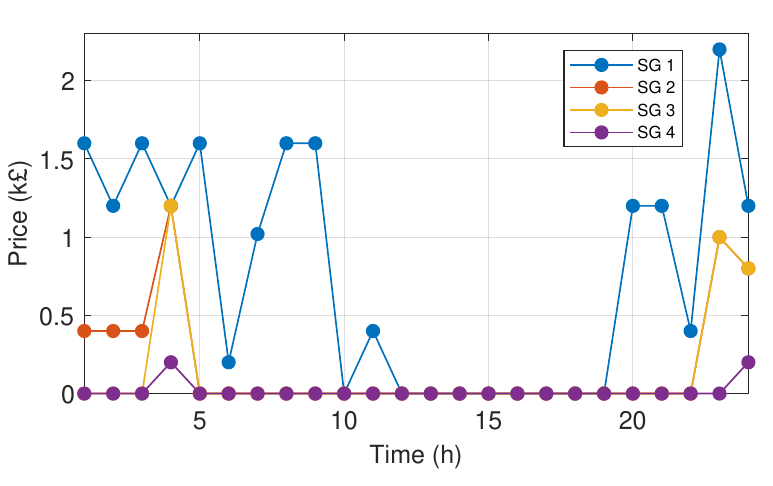}
    \caption{\label{fig:P3_hour}SG commitment prices during one-day operation.}
    \vspace{-0.4cm}
\end{figure} 
Furthermore, the SG commitment prices at each hour during one-day operation is depicted in Fig.~\ref{fig:P3_hour}, where only the contribution to the bus of IBR 1 is shown due to the space limitation and only the first four SGs are plotted for clarity. It is clear from the figure that the SGs that have a larger capacity and are closer to IBR 1 have a higher value in the SCC improvement. Similar to the dispatchable method, in the hours when the SCC is sufficient, the commitment price for the SGs is also zero.

\section{Conclusion} \label{sec:5}
An SCC pricing model in high IBR-penetrated systems is developed in this work by incorporating the SCC constraints into the system scheduling problem. Three different pricing schemes dealing with the nonconvex nature of the SCC-constrained UC problem are analyzed and compared. The results obtained based on the modified IEEE 39-bus system show that the SCC prices with the dispatchable method and marginal unit prices of the SGs are sensitive to the locations and system operating conditions. The restricted method is demonstrated to be not suitable for the SCC pricing due to the coupling with other constraints. The future work involves investigating the SCC and unit prices considering the interactions among different sinks and sources.


\vspace{-0.25cm}
\bibliographystyle{IEEEtran}
\bibliography{bibliography}





\end{document}


\begin{varwidth}{\linewidth}

\begin{tikzpicture}
\begin{axis}[
    ybar=0.8pt,
    bar width=.02cm,
    width=7.25cm,
    height=4cm,
    symbolic x coords={1,2,3,4,5,6,7,8,9,10,11,12,13,14,15,16,17,18,19,20,21,22,23,24,25,26,27,28,29,30},
    enlarge x limits=0.03,
    xlabel={$\mathrm{Bus}$},
    ylabel={$\mathrm{SCC}\,\mathrm{[p.u.]}$},
    xtick={5,10,15,20,25,30},
    xmajorgrids=true,
    ymajorgrids=true,
    legend style={at={(0.635,0.982)}, anchor=north,legend columns=-1}, legend cell align={left},
    grid style=dashed,nodes={scale=0.75, transform shape}]
\footnotesize
\addplot[
    thick,
    color=pBlue,
    fill=pBlue, 
    ]
    table {data/WL_SI/data1.txt};    
    \addlegendentry{\footnotesize $\rho_w$ = 0}        
\addplot[
    thick,
    color=pRed,
    fill=pRed, 
    ]
    table {data/WL_SI/data2.txt};        
    \addlegendentry{\footnotesize $\rho_w$ = 0.4}   
    
\addplot[
    thick,
    color=pYellow,
    fill=pYellow, 
    ]
    table {data/WL_SI/data3.txt};        
    \addlegendentry{\footnotesize $\rho_w$ = 0.8}        

\end{axis}

\end{tikzpicture} 

\end{varwidth}